\documentclass{emulateapj}

\shorttitle{SDSS QUASAR LENS SEARCH. II.}
\shortauthors{INADA ET AL.}

\begin{document}

\title{The Sloan Digital Sky Survey Quasar Lens Search. II. \\
  Statistical Lens Sample from the Third Data Release}

\author{
Naohisa Inada,\altaffilmark{1,2} 
Masamune Oguri,\altaffilmark{3,4} 
Robert H. Becker,\altaffilmark{5,6} 
Min-Su Shin,\altaffilmark{4} 
Gordon T. Richards,\altaffilmark{7}\\
Joseph F. Hennawi,\altaffilmark{8}
Richard L. White,\altaffilmark{9}
Bartosz Pindor,\altaffilmark{10}
Michael A. Strauss,\altaffilmark{4} \\
Christopher S. Kochanek,\altaffilmark{11} 
David E. Johnston,\altaffilmark{12,13} 
Michael D. Gregg,\altaffilmark{5,6} 
Issha Kayo,\altaffilmark{14} \\
Daniel Eisenstein,\altaffilmark{15} 
Patrick B. Hall,\altaffilmark{16} 
Francisco J. Castander,\altaffilmark{17} 
Alejandro Clocchiatti,\altaffilmark{18} \\
Scott F. Anderson,\altaffilmark{19} 
Donald P. Schneider,\altaffilmark{20} 
Donald G. York,\altaffilmark{21,22} 
Robert Lupton,\altaffilmark{4} \\
Kuenley Chiu,\altaffilmark{23} 
Yozo Kawano,\altaffilmark{14} 
Ryan Scranton,\altaffilmark{24} 
Joshua A. Frieman,\altaffilmark{22,25,26} 
Charles R. Keeton,\altaffilmark{27} \\
Tomoki Morokuma,\altaffilmark{28} 
Hans-Walter Rix,\altaffilmark{29} 
Edwin L. Turner,\altaffilmark{4} 
Scott Burles,\altaffilmark{30,31} \\
Robert J. Brunner,\altaffilmark{32} 
Erin Scott Sheldon,\altaffilmark{33}
Neta A. Bahcall,\altaffilmark{4} and
Masataka Fukugita\altaffilmark{34}
}

\altaffiltext{1}{Cosmic Radiation Laboratory, RIKEN (The Physical and
                 Chemical Research Organization), 2-1 Hirosawa, Wako,
                 Saitama 351-0198, Japan.} 
\altaffiltext{2}{Institute of Astronomy, Faculty of Science, University
                 of Tokyo, 2-21-1 Osawa, Mitaka, Tokyo 181-0015, Japan.} 
\altaffiltext{3}{Kavli Institute for Particle Astrophysics and
                 Cosmology, Stanford University, 2575 Sand Hill Road,
                 Menlo Park, CA 94025.} 
\altaffiltext{4}{Princeton University Observatory, Peyton Hall,
                 Princeton, NJ 08544.}  
\altaffiltext{5}{IGPP-LLNL, L-413, 7000 East Avenue, Livermore, CA 94550.}
\altaffiltext{6}{Department of Physics, University of California at
                 Davis, 1 Shields Avenue, Davis, CA 95616.}  
\altaffiltext{7}{Department of Physics, Drexel University, 3141
                 Chestnut Street,  Philadelphia, PA 19104.}
\altaffiltext{8}{Department of Astronomy, University of California at
                 Berkeley, 601 Campbell Hall, Berkeley, CA
                 94720-3411.}
\altaffiltext{9}{Space Telescope Science Institute, 3700 San Martin
                 Drive, Baltimore, MD 21218.} 
\altaffiltext{10}{Space Research Centre, University of Leicester.}
\altaffiltext{11}{Department of Astronomy, The Ohio State University, 
                  Columbus, OH 43210.}                 
\altaffiltext{12}{Jet Propulsion Laboratory, 4800 Oak Grove Drive,  
                 Pasadena CA, 91109.}
\altaffiltext{13}{California Institute of Technology, 1200 East  
                 California Blvd, Pasadena, CA 91125.}
\altaffiltext{14}{Department of Physics and Astrophysics, Nagoya
                  University, Chikusa-ku, Nagoya 464-8062, Japan.}
\altaffiltext{15}{Steward Observatory, University of Arizona, 933 North 
                  Cherry Avenue, Tucson, AZ 85721.}
\altaffiltext{16}{Department of Physics and Astronomy, York University,
                  4700 Keele Street, Toronto, Ontario, M3J 1P3, Canada.}
\altaffiltext{17}{Institut d'Estudis Espacials de Catalunya/CSIC,
                  Gran Capita 2-4, 08034 Barcelona, Spain.}
\altaffiltext{18}{Departamento de Astronom\'{i}a y Astrof\'{i}sica,
                 Pontificia Universidad Cat\'{o}lica de Chile, Casilla
                 306, Santiago 22, Chile.}
\altaffiltext{19}{Astronomy Department, Box 351580, University of Washington, 
                  Seattle, WA 98195.}
\altaffiltext{20}{Department of Astronomy and Astrophysics, The
                  Pennsylvania State University, 525 Davey Laboratory, 
                  University Park, PA 16802.}   
\altaffiltext{21}{Department of Astronomy and Astrophysics, The University 
                  of Chicago, 5640 South Ellis Avenue, Chicago, IL 60637.}
\altaffiltext{22}{Enrico Fermi Institute, The University of Chicago,
                  5640 South Ellis Avenue, Chicago, IL 60637.}
\altaffiltext{23}{School of Physics, University of Exeter, Stocker Road, 
                  Exeter EX4 4QL, UK.}
\altaffiltext{24}{University of Pittsburgh, Department of Physics and
                 Astronomy, 3941 O'Hara Street, Pittsburgh, PA 15260.}
\altaffiltext{25}{Center for Particle Astrophysics, Fermilab, P.O. Box 500, 
                  Batavia, IL 60510.}
\altaffiltext{26}{Kavli Institute for Cosmological Physics, University of 
                  Chicago, Chicago, IL 60637.}
\altaffiltext{27}{Department of Physics and Astronomy, Rutgers University, 
                  Piscataway, NJ 08854.}
\altaffiltext{28}{National Astronomical Observatory, 2-21-1 Osawa, Mitaka, 
                  Tokyo 181-8588, Japan.}
\altaffiltext{29}{Max Planck Institute for Astronomy, Koenigsstuhl 17, 
                  69117 Heidelberg, Germany.}
\altaffiltext{30}{Department of Physics, Massachusetts Institute of
                 Technology, 77 Massachusetts Avenue, Cambridge, MA 02139.}
\altaffiltext{31}{Kavli Institute for Astrophysics and Space Research,
                  Massachusetts Institute of Technology, Cambridge, MA 02139.}
\altaffiltext{32}{Department of Astronomy, University of Illinois, 1002 West 
                  Green Street, Urbana, IL 61801.}
\altaffiltext{33}{Center for Cosmology and Particle Physics, 
                  Department of Physics, New York University, 4
                 Washington Place, New York, NY 10003.}   
\altaffiltext{34}{Institute for Cosmic Ray Research, University of Tokyo, 
                  5-1-5 Kashiwa, Kashiwa, Chiba 277-8582, Japan.} 

\begin{abstract}
We report the first results of our systematic search for strongly
lensed quasars using the spectroscopically confirmed quasars in the
Sloan Digital Sky Survey (SDSS). Among 46,420 quasars from the 
SDSS Data Release 3 ($\sim$4188~deg$^2$), we select a subsample of
22,683 quasars that are located at redshifts between 0.6 and 2.2 and are
brighter than the Galactic extinction corrected $i$-band magnitude of
19.1. We identify 220 lens candidates from the quasar subsample, for
which we conduct extensive and systematic follow-up observations in
optical and near-infrared wavebands, in order to construct a complete
lensed quasar sample at image separations between $1''$ and $20''$ and
flux ratios of faint to bright lensed images larger than
$10^{-0.5}$. We construct a statistical sample of 11 lensed quasars. Ten
of these are galaxy-scale lenses with small image separations ($\sim
1''-2''$) and one is a large separation ($15''$) system which is
produced by a massive cluster of galaxies, representing the first
statistical sample of lensed quasars including both galaxy- and
cluster-scale lenses. The Data Release 3 spectroscopic
quasars contain an additional 11 lensed quasars outside the
statistical sample. 
\end{abstract}


\section{Introduction}\label{sec:intro}

Large systematic surveys of gravitationally lensed quasars are
essential for various scientific applications, as shown in 
a recent review by \citet{kochanek06}. For example, lensing
probabilities from large homogeneous surveys, which can be estimated
from the number of lenses in a statistically well-defined sample of
quasars, offer a probe of cosmological parameters. The largest
existing survey, the Cosmic-Lens All Sky Survey
\citep[CLASS;][]{myers03,browne03} contains a total of 22 lenses
discovered from high-resolution imaging of over 16,000 flat spectrum
radio sources. A subset of 13 lenses from 8,958 radio sources
constitutes a statistically well-defined lens sample which has been
used to study cosmological parameters as well as the structure of lens
galaxies \citep[e.g.,][]{rusin01,chae02,chen04,chae06,mitchell05}. 
However, a drawback of radio lens surveys like CLASS is that the
redshift distribution of the source population, which is a key
component for statistical analyses, is poorly constrained
\citep[e.g.,][]{munoz03}. Thus we will benefit from complementary
optical lens samples for which source populations are better
understood, although the effects of absorption and emission by
the lensing galaxies are larger in the optical than the radio
\citep[e.g.,][]{falco99}. The largest existing statistical sample of
optical lensed quasars is the Hubble Space Telescope snapshot survey 
\citep{maoz93}. It contains only five lenses selected from 502 bright 
high-redshift quasars, indicating the need for much larger optical
lens samples. 

Larger statistical lens samples will also allow the study of the
formation of structure. Quasars can be lensed by structures on scales
from individual galaxies, through groups, to clusters, and therefore
the image separation distribution of strongly lensed quasars from
small to large separations directly reflects the hierarchical 
structure formation and the effects of cooling the baryons
\citep[e.g.,][]{kochanek01,oguri06a}. Unfortunately, the probability
of quasars strongly lensed by clusters is 1--2 orders of magnitudes
smaller than that by galaxies, so we need large homogeneous surveys to
study the full image separation distribution. Indeed, despite
searching for them explicitly, no cluster-scale lens was discovered in
the CLASS \citep{phillips01}. 

The main purpose of the Sloan Digital Sky Survey Quasar Lens Search 
\citep[SQLS;][hereafter Paper I]{oguri06b} is to construct a large
sample of lensed quasars in the optical. It is made possible by the
large spectroscopic quasar catalog obtained from the data of the Sloan
Digital Sky Survey \citep[SDSS;][]{york00}. Lens candidates are
selected morphologically among the spectroscopically confirmed SDSS
quasars. Additional lens candidates are selected by looking for
companion objects to the SDSS quasars that have similar colors. Our
selection algorithms have been tested against simulated SDSS images;
this allows accurate quantification of the selection function (see Paper 
I). A number of previous strong lens discoveries \citep[][and 
references therein]{inada07a} indicate the effectiveness of our
candidate selection algorithm. 

In this paper, we present the statistical sample of strongly lensed 
quasars, constructed from the SDSS Data Release 3 
\citep[DR3;][]{abazajian05} spectroscopic quasar catalog
\citep{schneider05}. Lens candidates are selected according to the
algorithms presented in Paper I. We conduct extensive follow-up
observations for these candidates with various facilities in order to
test the hypothesis that they are lensed, and to make a complete
lens sample. Cosmological constraints from this statistical sample will
be reported in Paper III of this series \citep{oguri07a}. 

The structure of this paper is as follows. We describe the construction
of the source quasar sample in \S \ref{sec:source}, and the selection of
lens candidates in \S \ref{sec:cand}. Results of our follow-up
observations for the candidates are summarized in \S \ref{sec:obs}. 
Section \ref{sec:dr3_lens} lists lensed quasars in the statistical
quasar subsample as well as those included in the DR3 quasar catalog. 
Finally we summarize our results in \S \ref{sec:conc}. 

\section{Source Quasar Sample}\label{sec:source}

\begin{figure*}
\epsscale{.9}
\plotone{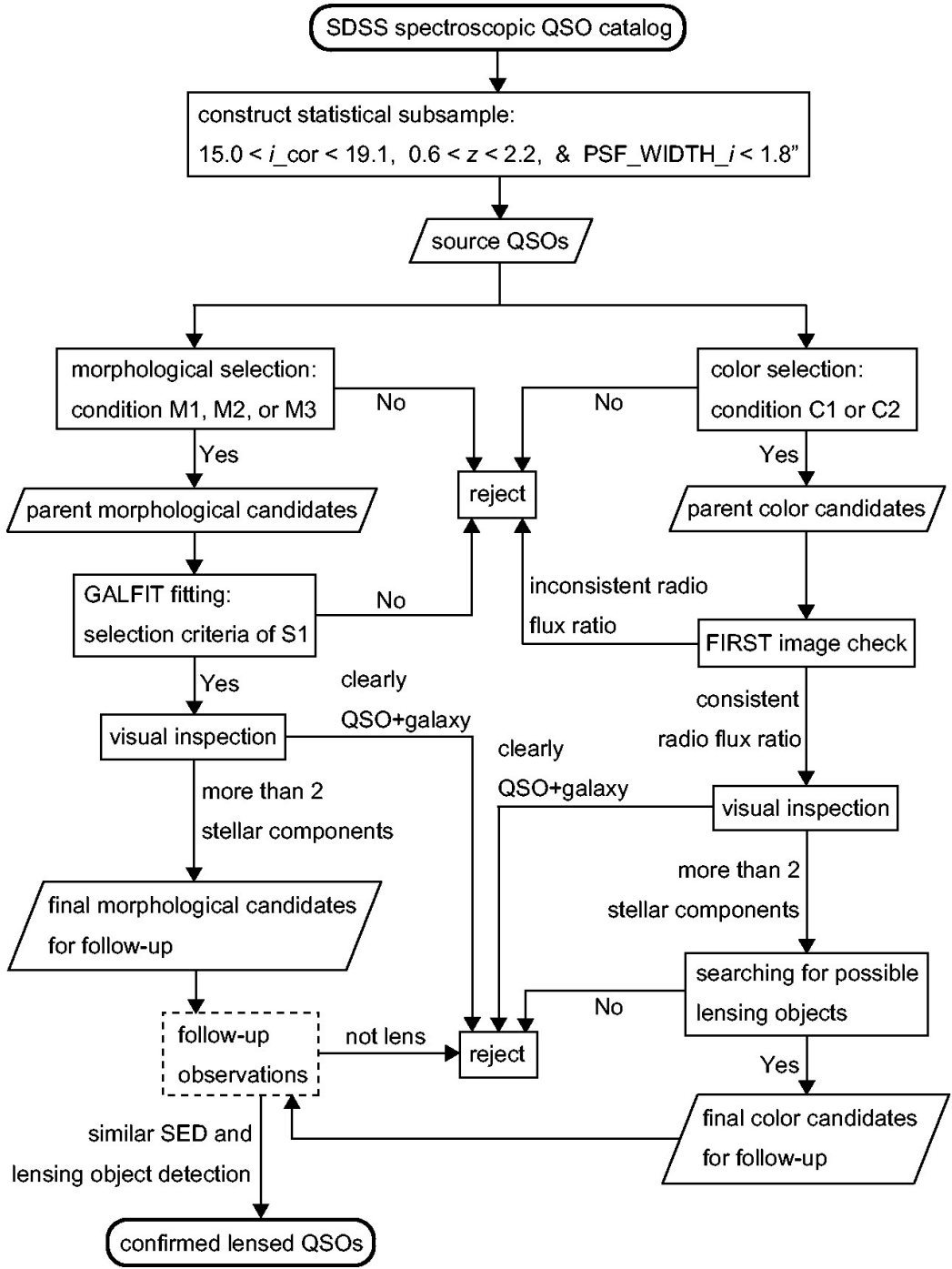}
\caption{Flowchart of the candidate selection procedure of the SQLS.
  First we construct a statistical subsample of quasars (source QSOs) 
  from the SDSS spectroscopic quasar catalog. The specific selection 
  criteria (M1--M3, C1--C2, and S1) are given in Paper I. The details 
  of the additional selection criteria are described in \S \ref{sec:cand}. 
  Table \ref{tab:cand} presents the numbers of the source quasars, 
  parent candidates, objects rejected at each step, and final follow-up 
  candidates.
\label{fig:flowchart}}
\end{figure*}

The SDSS is a combination of photometric and spectroscopic surveys of
a quarter of the entire sky \citep{york00}. It uses a dedicated
wide-field ($3^{\circ}$ field of view) 2.5-m telescope \citep{gunn06}
at the Apache Point Observatory in New Mexico, USA. Photometric
observations \citep*{gunn98,tucker06} consist of imaging in five broad
band filters \citep{fukugita96}. The data are processed and analyzed
automatically by the photometric pipeline \citep{lupton01,lupton07}. 
Targets for spectroscopy are selected according to selection algorithms
applied to the imaging data (see \citet{richards02} for the quasar
target selection algorithm). Spectra of these targets are obtained
with a multi-fiber spectrograph (wavelength range between 3800{\,\AA}
and 9200{\,\AA} at a resolution of R$\sim$1800). The astrometric
accuracy of the imaging data is better than about $0\farcs1$ rms per
coordinate \citep{pier03} and photometric zeropoint errors are less than
about 0.03 magnitude over the entire survey area
\citep{hogg01,smith02,ivezic04,padmanabhan07}. The data release papers
\citep{stoughton02,abazajian03,abazajian04,abazajian05,adelman06,adelman07}
describe the contents of the SDSS data releases. 

We start with a sample of 46,420 spectroscopically confirmed quasars in
the SDSS DR3 quasar catalog \citep{schneider05}. The area covered by the
spectroscopy is 4188~${\rm deg^2}$. The sample contains quasars from
$z=0.08$ to $5.41$ with a median redshift of 1.47. This is {\it not} a
well-defined quasar sample for lens surveys, as it includes objects
selected with a wide variety of techniques. For example, when high
redshift quasar candidates ($z>2.2$) are targeted for SDSS spectroscopy,
they are required to be point sources, leading to a strong bias against
selecting small separation lenses. We focus on the low redshift
($z<2.2$) and bright ($i_{\rm cor}<19.1$; here $i_{\rm cor}$ is the
Point Spread Function magnitude corrected for Galactic extinction from
the maps of \citet{schlegel98}) quasars of the main quasar sample,
which are known to have high completeness regardless of whether they
are resolved or unresolved \citep{vandenberk05,richards06}. Thus the
quasars with $z<2.2$ and $i_{\rm cor}<19.1$ should have no explicit
biases against gravitational lenses. We further restrict the redshift
range to $0.6<z<2.2$ to eliminate lower redshift, intrinsically extended
quasars, and exclude quasars with SDSS images of poor seeing ({\tt
PSF\_WIDTH}$>1\farcs8$) in which the identification of close lens pairs
is difficult. Paper I discusses the selection of the source quasars in
greater detail. These criteria produce a subsample of 22,683
quasars\footnote{This includes one quasar (SDSS~J094222.89+102025.3),
which we missed in Paper I. See
http://www-utap.phys.s.u-tokyo.ac.jp/~sdss/sqls/ for the redshift and
$i_{\rm cor}$ distributions of our quasar subsample.} suitable for
lens statistics. From this quasar subsample, we construct a
statistical sample of lensed quasars. 

\section{Lens Candidate Selection}\label{sec:cand}

We illustrate the SQLS candidate selection procedure in Figure
\ref{fig:flowchart}. As discussed in Paper I, we use two different
selection methods (morphological and color selection), in order to
identify both galaxy- and cluster-scale lens candidates. For
candidates selected by each approach, we apply several additional
selection criteria to construct a final lens candidate sample
appropriate for detailed follow-up on other facilities. We explain 
these additional criteria for morphological candidates 
in \S~\ref{sec:morp} and those for color candidates in 
\S~\ref{sec:col}. These two selection algorithms are not exclusive 
with each other, since the (deblended) quasar component of a color 
candidate could be a morphological candidate. The numbers of candidates 
selected/removed by each approach are summarized in Table \ref{tab:cand}. 
We finally identify 220 lensed quasar candidates for follow-up from the 
22,683 source quasars. 

\begin{deluxetable}{lr}
\tablecaption{NUMBERS OF CANDIDATES\label{tab:cand}}
\tablewidth{0pt}
\tablehead{ 
\colhead{} & 
\colhead{Number} }
\startdata
Source quasars & 22,683 \\ \hline
Parent morphological candidates                 &    649  \\  
Rejected by GALFIT fitting                      & $-$552  \\
Rejected by visual inspection                   &   $-$7  \\
\hline
Final morphological candidates for follow-up    &  90  \\   
\hline
Parent color candidates                         &    227  \\  
Rejected by FIRST image check                   &  $-$16  \\
Rejected by visual inspection                   &  $-$16  \\
Rejected by searching for possible lensing objects  &  $-$63  \\
\hline
Final color candidates for follow-up            & 132  \\   
\hline\hline
Final total candidates for follow-up            & 220  \\
\enddata
\tablecomments{Two candidates are selected by both the morphological 
 and color selection algorithms. }
\end{deluxetable}

\subsection{Morphological Selection}\label{sec:morp}

The morphological selection algorithm is intended to discover
galaxy-scale ($\theta\lesssim 2\farcs5$) lensed quasars that the SDSS 
photometric pipeline did not deblend into multiple components. In Paper
I, we showed that such lens candidates can be identified by searching
for quasars that are not well fit by the PSF in each SDSS field. Thus
we select the galaxy-scale lens candidates based on the goodness of fit
of a quasar image to the PSF model in each field ({\tt star\_L}). 
Different criteria of {\tt star\_L} are used according to the object
classification ({\tt objc\_type}), which is set by the difference
between the PSF and model magnitudes in the SDSS data. The specific
criteria (M1, M2, or M3) are given in Paper I. These ``morphological
selection'' criteria identify 649 quasars ($\sim 3\%$ of the source 
quasars) as lens candidates. 

As discussed in Paper I, a significant fraction of the candidates selected 
by the algorithm are single quasars. In addition, the candidates contain 
superpositions of quasars with foreground galaxies or stars. Therefore, 
we use two additional criteria described below in order to exclude most 
of these false positives before performing any subsequent observations. 
In the first criterion, we fit the SDSS $u$- and $i$-band images of each 
candidate with two PSFs using the GALFIT software \citep{peng02}. When 
a single quasar is fitted with two PSFs, the result tends to be unusual 
in the sense that the fitted two PSFs have a very small separation ($\ll
1''$) and similar magnitudes, or a moderate separation ($\gtrsim 1''$)
and very different magnitudes. In addition, when a single quasar plus
star/galaxy system is fitted with two PSFs, the results from the SDSS
$u$- and $i$-band images tend to be very different because a quasar at
$0.6<z<2.2$ is usually much bluer than either a star or a galaxy. In
particular, such quasar plus star/galaxy systems may result in very large
$u$-band flux ratios since the galaxy/star component tends to be very
faint in $u$-band. Thus we can eliminate a significant fraction of these
false positives by making cuts to the fitted separations and flux ratios
(see Paper I for selection criterion S1). This procedure (``GALFIT
fitting'') successfully removes $\sim 85\%$ of the candidates. Most of
the rejections are single isolated quasars, but as expected, they also
include several quasar-star and quasar-galaxy pairs. 

For the second criterion, we inspect the SDSS images of the 
remaining candidates and eliminate those which appear to be chance
superpositions of a quasar and a galaxy (``visual inspection'').
This visual inspection eliminates about 10\% of the candidates left
after the GALFIT modeling and yields 90 morphological (galaxy-scale)
candidates that require further investigation.

\subsection{Color Selection}\label{sec:col}

Larger separation ($\theta\gtrsim 2\farcs5$) lenses created by groups
or clusters of galaxies are accurately deblended by the SDSS photometric
pipeline, and therefore can be selected by comparing the colors of each
quasar to those of nearby objects (``color selection''). The idea is
similar to the \citet{hennawi06} approach for finding binary quasars in
the SDSS, but we modify the color selection criteria to allow for
differential extinction between lensed images. These criteria are 
discussed in detail in Paper I, but we have slightly broadened the
limits to include image separations of $\theta<20\farcs1$ and $i$-band
magnitude differences of $\Delta i<1.3$, considering the typical
uncertainties in the quasar positions ($0\farcs1$) and magnitudes
($0.05$) in the SDSS data. We identified 227 quasar pairs based on the
criteria. 

We first test the lensing hypothesis for these candidates by comparing
the radio flux ratios from the Faint Images of the Radio Sky at Twenty
centimeters survey \citep[FIRST;][]{becker95} with the optical flux
ratios (``FIRST image check''). We eliminate 10\% of the
candidates with separations larger than $6''$ (set by the $\sim
5''$ resolution of the FIRST) in which one component is radio loud with 
a flux well above the FIRST survey limits and the other is not. Such
pairs are either binary quasars or quasar-star pairs \citep{kochanek99}.

Next we check the SDSS images of the candidates (``visual inspection''). 
Since we select both point sources and extended sources as the nearby 
objects in the color selection stage, quasar plus galaxy systems 
are sometimes selected as candidates. The purpose of visual inspection
in the color selection algorithm is to eliminate the candidates whose 
companion objects to the SDSS quasars clearly appear to be galaxies. 
At this step an additional $\sim 10\%$ of the candidates are excluded. 

\begin{figure}
\epsscale{.95}
\plotone{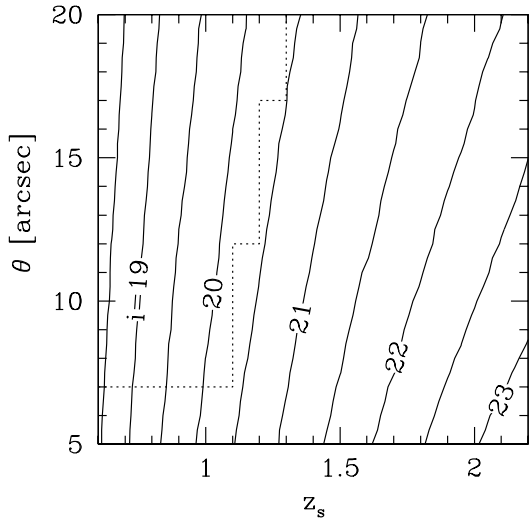}
\caption{The $i$-band magnitude limit of the lensing objects (defined 
  such that 99\% of simulated lenses are caused by galaxies brighter 
  than the limit) in the $z_s$-$\theta$ plane, where $z_s$ denotes the 
  source (quasar) redshift. The limit is computed using the halo model 
  of \citet{oguri06a}. The dotted line indicates the limits of the 
  source redshift and image separation (eq. [\ref{eq:zscut}]): 
  For candidates which lie in the region of eq. [\ref{eq:zscut}], 
  we search for possible lensing galaxies in the SDSS $i$-band images 
  and reject the candidates if no galaxy is seen among the components. 
  See text for more details. 
\label{fig:mag_cont}}
\end{figure}

In addition, for low-redshift candidates we can search for possible
lensing galaxies in the SDSS images (``searching for possible lensing
objects''). To determine the criteria, we compute the expected
magnitudes of the lensing objects using the halo model of
\citet{oguri06a}. Specifically, we define the magnitude limit of the
lensing galaxy (or the central galaxy of the lensing cluster for a
cluster-scale lens candidate) such that 99\% of simulations have a 
lensing galaxy brighter than that magnitude limit, and compute the 
magnitude limit as a function of the image separation and source 
(quasar) redshift. In the model, halos are linked to galaxies by 
adopting a universal scaling relation (with scatter) between masses 
of halos and luminosities of galaxies. We ignore the redshift 
evolution of the mass-luminosity relation, which provides conservative
estimates of the minimum luminosity since standard passive evolution
predicts that galaxies were brighter in the past. The luminosity is
converted to observed magnitude using the K-correction for elliptical
galaxies in \citet{fukugita95}. Figure~\ref{fig:mag_cont} shows the
magnitude limit as a function of image separations and quasar
redshifts. We check for a lensing galaxy in the SDSS image when its
expected magnitude is brighter than $i=20.5$ (corresponding to
$\sim 0.5L_*$ for a typical lens redshift); given the magnitude limit of
the SDSS images, $i_{\rm lim}\sim 21.5$, the choice should be very
conservative. In practice, we examine candidates with low redshifts and
large separations defined by the region shown in Figure~\ref{fig:mag_cont}: 
\begin{eqnarray}
&&z_s<1.1 \;\;\;\;\; \mbox{for $7''<\theta<12''$},\nonumber\\
&&z_s<1.2 \;\;\;\;\; \mbox{for $12''<\theta<17''$},\nonumber\\
&&z_s<1.3 \;\;\;\;\; \mbox{for $17''<\theta$},
\label{eq:zscut}
\end{eqnarray}
where $z_s$ is the redshift of the source quasar. The candidate which 
lies in the region and lacks a possible lensing object in the SDSS image
is rejected. Roughly 30\% of the candidates fail this test, leaving 132
color (larger-separation) candidates. These 132 objects constitute the
final color candidates for additional investigation. Among these, 
two candidates were also selected by the morphological algorithm, which
is consistent with our simulation that predicts $\sim 10\%$ of objects
with image separations of $1\farcs5\lesssim\theta\lesssim3\farcs0$ are
selected by both algorithms (see Paper I). Thus the total number of
candidates for follow-up is 220.  

\section{Follow-up Observations}\label{sec:obs}

The final morphological and color candidates and the summary of
their observations are shown in Tables \ref{tab:can_mor}
and \ref{tab:can_col}, respectively. In this section we describe the
follow-up observations and how we decide on the lens nature of each
candidate. We also note several interesting objects discovered in the
course of our lens search. 

\subsection{Basic Strategy}\label{sec:strategy}

Before conducting any subsequent observations, we first check if the
candidates have been studied before with the NASA/IPAC Extragalactic
Database (NED). Three of the 220 candidates are previously known
gravitational lens systems for which no follow-up observations were
necessary. One system, SDSS~J133945.37+000946.1, turns out to be a
quasar pair at different redshifts \citep{croom04}.

The rest of the candidates are investigated to test their lensing
hypotheses. These observations consist of optical spectroscopy,
optical imaging, and near-infrared imaging, conducted at the following
facilities: the University of Hawaii 2.2-meter telescope (UH88), the
Astrophysical Research Consortium 3.5-meter telescope (ARC 3.5-m), the
Keck I and II telescopes, the United Kingdom Infra-Red Telescope
(UKIRT), the Subaru telescope, the Magellan Consortium's Walter Baade
6.5-m telescope (WB 6.5-m), the Hubble Space Telescope (HST), the MDM
2.4-meter telescope (MDM 2.4-m), the MMT Observatory, the European
Southern Observatory 3.6-meter telescope (ESO 3.6-m), the New
Technology Telescope (NTT), and the WIYN telescope. 

The SDSS images have moderate observing conditions (typically $\sim 
1\farcs3$ seeing) and a short exposure time (about 55~sec), making it 
difficult in many cases to determine whether a lensing galaxy is present. 
Therefore, for small separation lens candidates we usually start with
acquiring deeper optical ($i$ or $I$) or infrared ($H$ or $K$) images
under good seeing conditions ($\sim 0\farcs5-1\farcs0$). Some of the
candidates turn out to be single quasars or quasar-galaxy pairs, and
therefore they are rejected rather easily. If the candidates consist 
of two (or more) stellar components, we take optical and/or infrared 
images that are deep enough to locate lensing galaxies. The typical 
magnitude limits for extended objects are $I \sim 23.0$, $H \sim 18.5$, 
and $K \sim 20.0$. Additional images with bluer filters may be taken in
order to better separate multiple components. Candidates that do not
exhibit any residuals after subtracting stellar components are 
rejected based on the absence of the lensing object. Some of our 
candidates are rejected simply by the fact that the separation of the 
two stellar components is smaller than $1''$, since we set the minimum 
image separation of our complete lens sample to be $1''$ (i.e., some of the 
``rejected'' candidates with $\theta<1''$ could in fact be gravitational 
lenses; see \S \ref{sec:note}.). If the data reveal stellar components
with similar colors and a possible lensing galaxy, we try to obtain
spectra of the multiple stellar components to determine whether their
spectral energy distributions (SEDs) are similar. We detect possible
lensing objects only for 9 candidates out of 81 morphological candidates
with deep images. Seven of them turn out to be new lenses from the SDSS
(see \S~\ref{sec:dr3_lens} and Tables \ref{tab:can_mor}), and the other
two are confirmed to be binary quasars (SDSS~J084710.40$-$001302.6 
and SDSS~J100859.55+035104.4; see \S~\ref{sec:note}). 
We note that in addition to the candidates with possible lensing
objects we conduct spectroscopic observations for any candidates which 
have stellar components with very similar colors in imaging data
(including the SDSS images) even if they are rejected based on the 
absence of lensing objects. We perform this spectroscopic observation 
in order to test the validity of the rejection criterion.
This includes SDSS~J093207.15+072251.3 (see \S~\ref{sec:note}) and
SDSS~J112012.11+671116.0 \citep{pindor06} that are confirmed to be
binary quasar pairs with indistinguishable redshifts.

Our strategy for follow-up of large separation lens candidates is similar
to the above process. We either acquire spectra of the two components to 
check their SEDs or deep optical/infrared images to search for any
possible lensing galaxies or clusters. Rejection based upon spectroscopic 
observations is straightforward when the candidates are quasar-star
pairs or quasar-quasar pairs at different redshifts. Some of the
candidates turn out to be binary quasars, as reported in
\citet{hennawi06}. In some cases candidates turn out to be quasar pairs
at quite similar redshifts. For these sources we use deep
optical/infrared follow-up images to search for any signature of a
lensing galaxy or a lensing cluster. As well as the small separation
lens candidates, we perform spectroscopic observations regardless of
the existences of the lensing objects particularly when the colors of
the two components are very similar. Most of them are quasar-quasar
pairs at different redshifts. Two of them, SDSS~J090955.54+580143.2 and
SDSS~J211102.60+105038.3, turn out to be quasar pairs with
indistinguishable redshifts, but are rejected based on the differences
of the SEDs (see \S~\ref{sec:note}). 

To summarize, we identify a candidate as a lens system when the following
three conditions are met: (i) the stellar components have the same
redshifts within the measurement uncertainty; (ii) their SEDs are 
reasonably similar; (iii) a galaxy or a cluster/group of galaxies is 
detected between the stellar components. When candidates have four 
stellar images, we do not always require conditions (i) and (ii),
because the object's lensing nature is obvious from its characteristic
image configuration. The fact that we use the existence of the
lensing object for the judgment suggests that our selection may be
against hypothetical dark lenses \citep[e.g.,][]{rusin02} that have
anomalously high mass-to-light ratios. However, since our candidates
with very similar colors tend to be rejected spectroscopically (see
above and \S~\ref{sec:note}) rather than the absence of lensing objects
alone, we believe that our follow-up strategy is reasonably effective in
locating such dark lenses as well. 

\begin{figure*}
\epsscale{.4}
\plotone{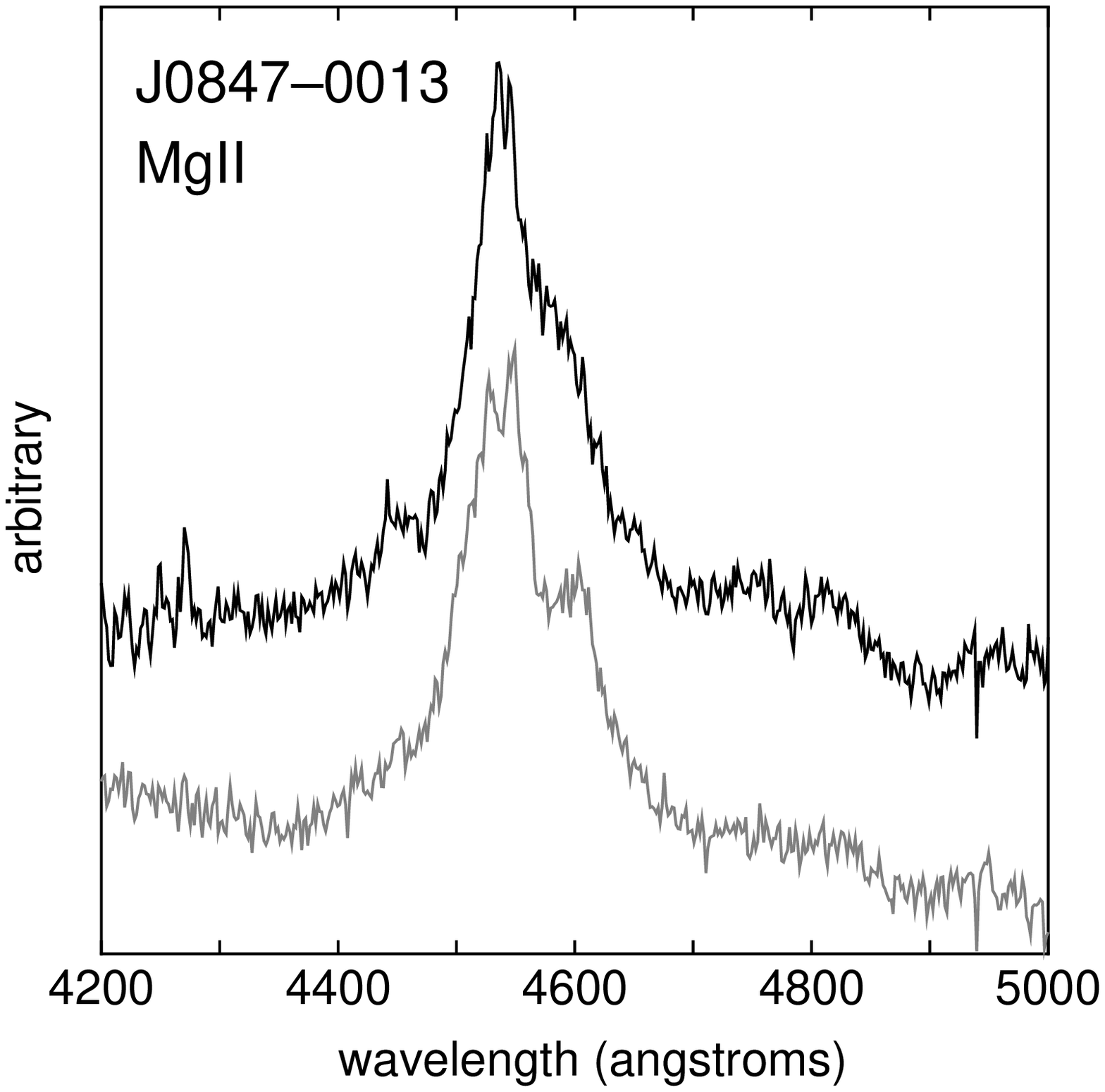}
\plotone{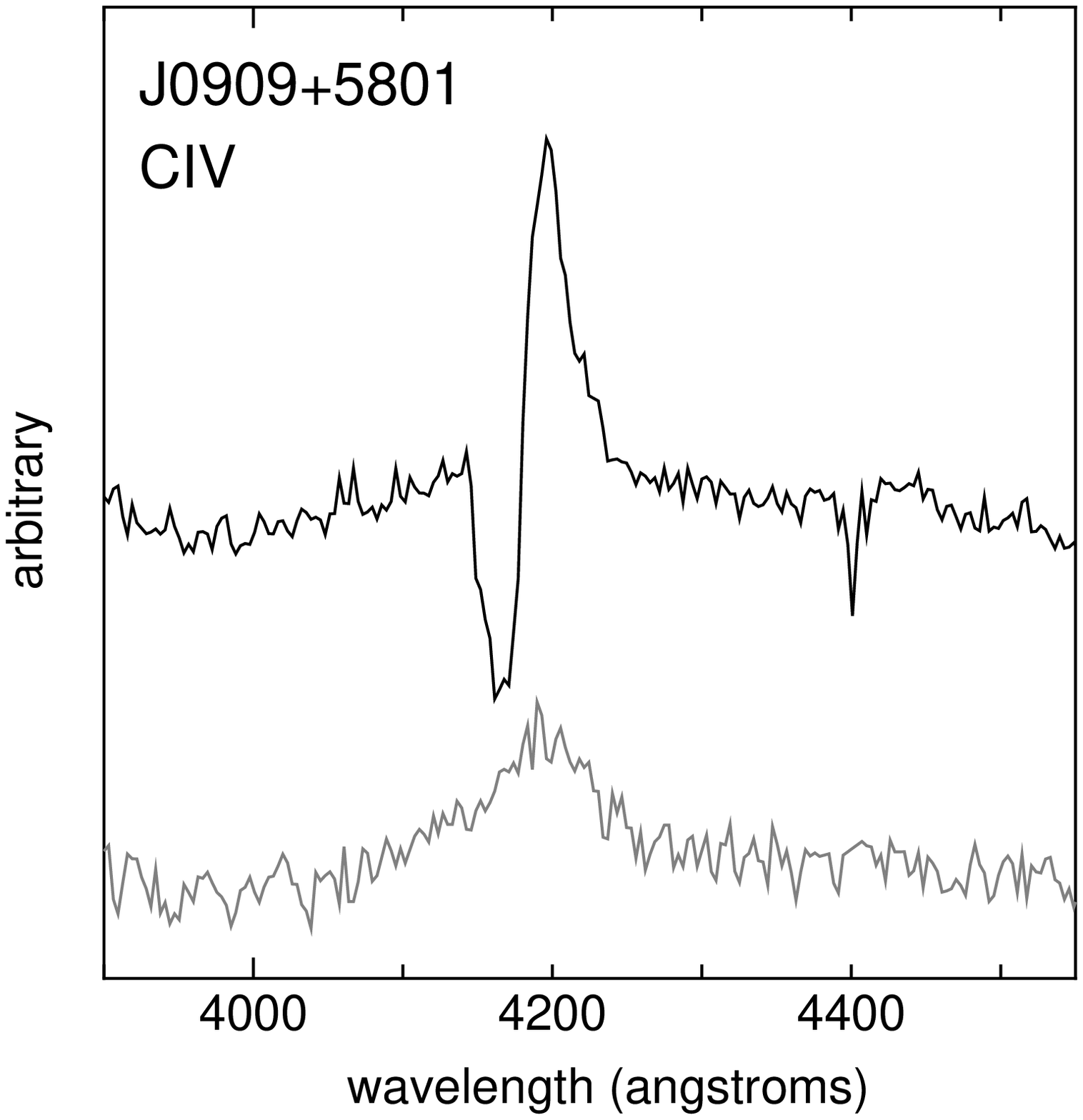}  \\
\plotone{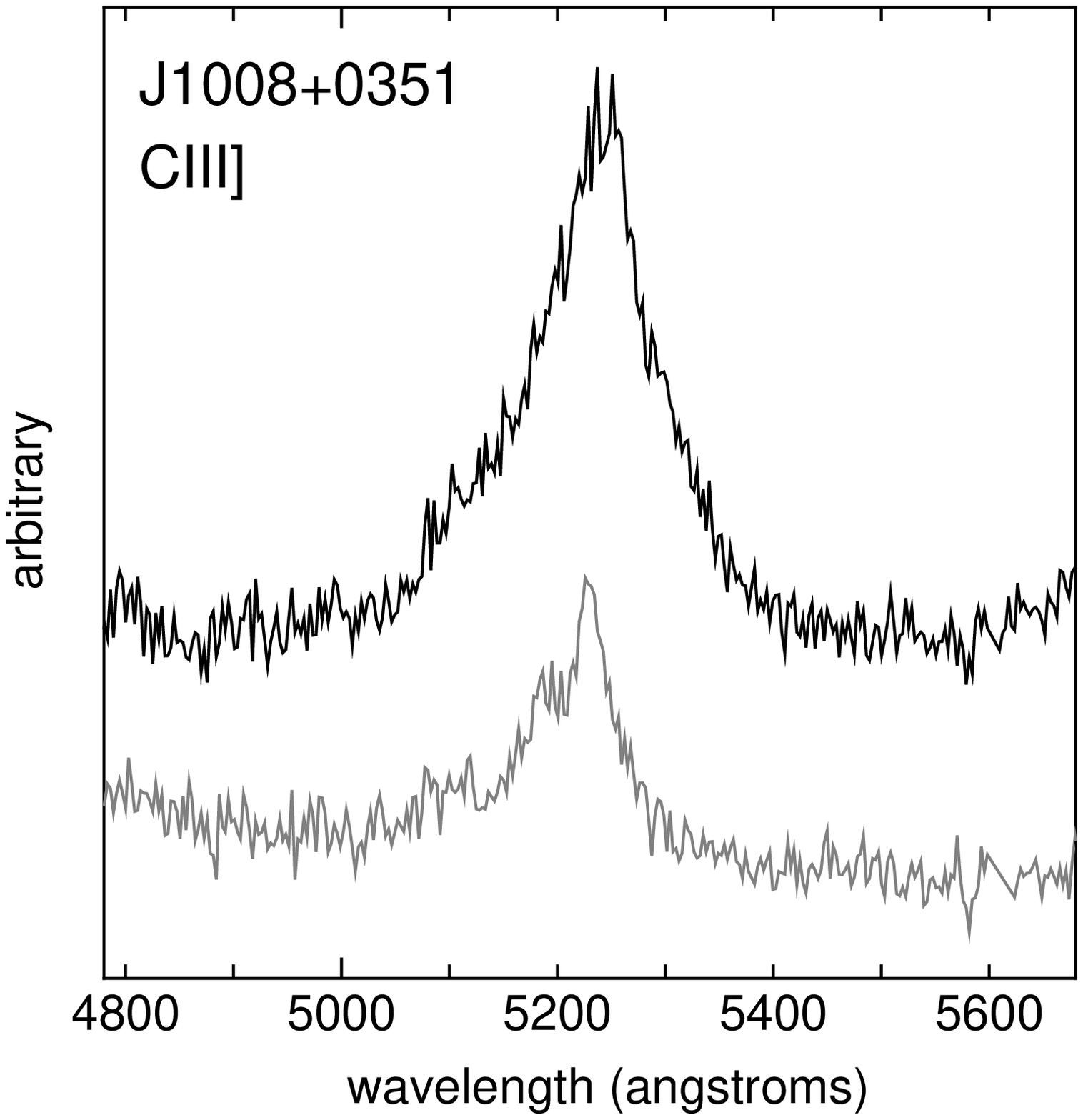}
\plotone{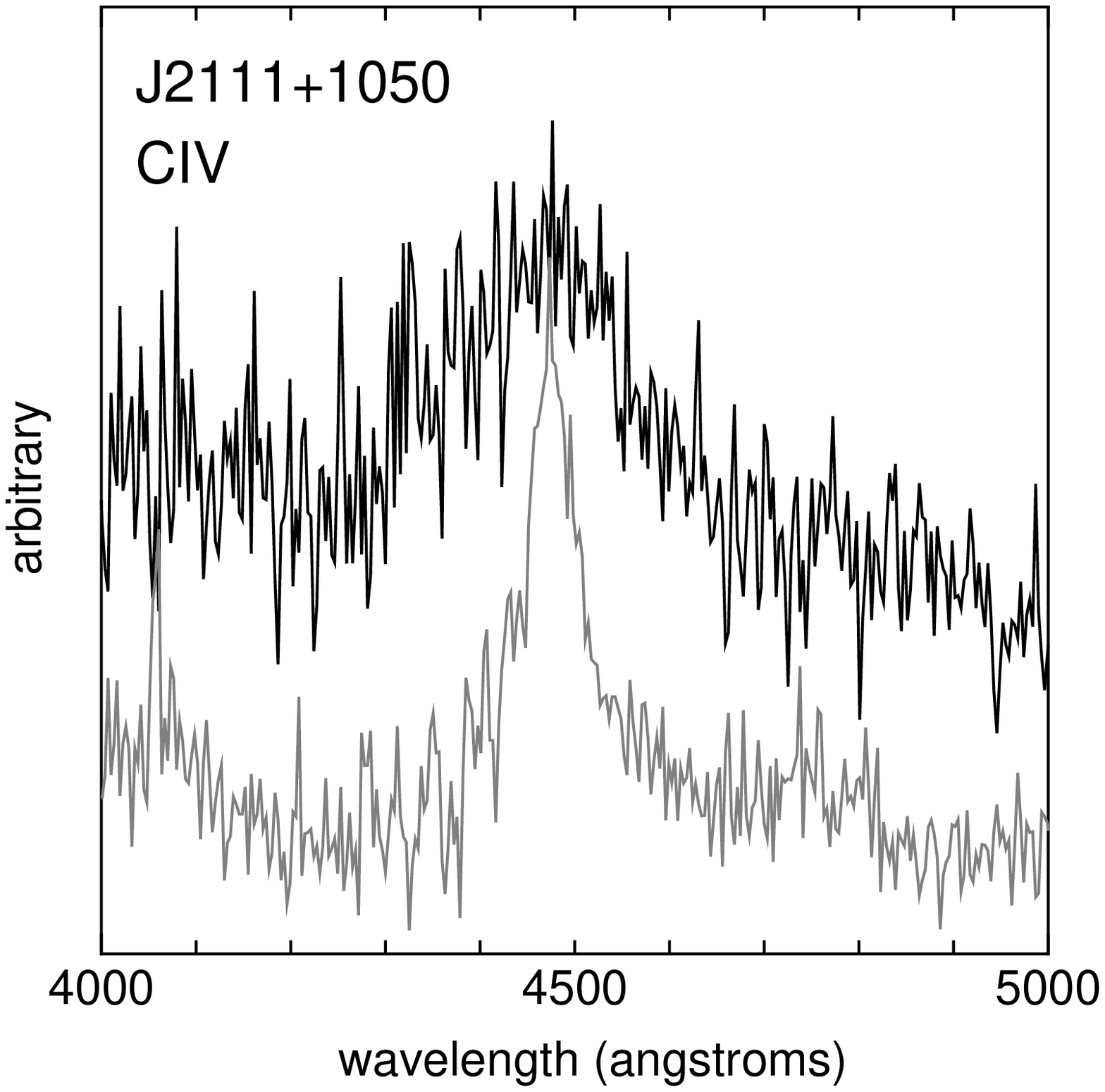}
\caption{
{\it Upper Left}: The \ion{Mg}{2} emission lines of the two quasar 
components of SDSS~J0847$-$0013 taken at the Keck telescope (spectral 
resolution of $R{\sim}1000$). The different shapes support the 
binary interpretation of this system. 
{\it Upper Right}: The \ion{C}{4} emission lines of the two quasar 
components of SDSS~J0909+5801 taken at the ARC 3.5-m telescope
($R{\sim}500$). The broad absorption line feature is seen only in the
 spectrum of the brighter component, which suggests that this is a
 binary quasar rather than a lens. 
{\it Lower Left}: The \ion{C}{3]} emission lines of the two quasar 
components of SDSS~J1008+0351 from the observation at the Subaru 
telescope ($R{\sim}500$). In addition to the different overall shapes, 
they show slightly different redshifts. 
{\it Lower Right}: The \ion{C}{4} emission lines of the two quasar 
components of SDSS~J2111+1050 observed at the ARC 3.5-m telescope
($R{\sim}500$). The different strengths of the emission lines,
together with the absence of any lensing objects in the deep $I$-band
image, support the binary interpretation.  
\label{fig:qq_pair_spec}}
\end{figure*}

\subsection{Notes on Individual Objects}\label{sec:note}

Below we note several interesting candidates which have not been
discussed elsewhere in the literature. In addition to the objects
discussed below, at least 30 objects out of our 220 candidates are
confirmed to be pairs of quasars; 9 out of the 30 pairs have already
been reported in \citet{hennawi06}. Furthermore, there are 9 candidates
whose image separations turned out to be $\theta<1''$ in the
follow-up studies. They include the known lensed quasar FBQ1633+3134
\citep{morgan01}. These systems are not included in the statistical lens
sample and therefore we did not perform any further spectroscopy or
deeper imaging. Although our current follow-up images do not show any
possible lensing objects for any of the candidates, lensing galaxies of
subarcsecond lenses are expected and observed to be faint. Very deep and
high resolution images are necessary to conclude whether they are lensed
quasars or not. See Tables \ref{tab:can_mor} and \ref{tab:can_col} for
more details of these objects. 

{\bf SDSS~J0847$-$0013:} This is a small separation ($\theta=1\farcs0$) 
lens candidate discovered by the morphological selection criterion. 
Spectroscopic observations conducted with LRIS at Keck revealed that 
this is a quasar pair with similar redshifts of $z=0.626$ and
$z=0.627$. The spectra of the two quasar components, however, have
different \ion{Mg}{2} emission line shapes (upper left panel of
Figure~\ref{fig:qq_pair_spec}), supporting the binary interpretation of
this system. The image taken with Tek2k at UH88 shows extended emission
in the vicinity of the brighter component. The spectrum of this emission
taken with LRIS at Keck indicates that this emission is due to the host
galaxy of the quasar \citep{gregg07}. 

{\bf SDSS~J0909+5801:} The spectroscopic observation at ARC 3.5-m
revealed that both the components separated by $\theta=8\farcs1$ are
quasars at same redshifts ($z=1.712$). However, we see a broad 
absorption line feature only in the \ion{C}{4} emission line of the 
brighter component (upper right panel of Figure~\ref{fig:qq_pair_spec}). 
Together with the absence of the lensing object in deep $i$-band image, 
we conclude that the system is a binary quasar rather than a lens. 

{\bf SDSS~J0932+0722:} A morphological candidate with an image separation 
of $\theta=1\farcs4$ found to be a pair of quasars at $z=1.994$ in Keck
ESI observations. None of
the deep optical and near-infrared images show any lensing galaxy 
between the two components, and the flux ratio of the two quasars 
in the optical are significantly different from that in the 
near-infrared (0.12 in $I$-band and 0.02 in $H$-band). Thus we regard 
this system as a binary quasar. 

{\bf SDSS~J1008+0351:} We detected possible extended emission between 
the stellar components in both deep optical and infrared
images. However, spectra of this small separation ($\theta=1\farcs1$)
candidate taken with FOCAS at Subaru suggest that this is a binary
quasar, because of the slightly different redshifts ($z=1.745$ and
$1.740$) and the different overall shapes of the spectra, as shown in
the lower left panel of Figure~\ref{fig:qq_pair_spec}. One
interpretation is that the extended emission is due to the host galaxy
of one of (or both of) the quasars. 

{\bf SDSS~J2111+1050:} This is a large separation ($\theta=9\farcs7$)
lens candidate from the color selected sample. From ARC 3.5-m
spectroscopy, the redshifts of the two stellar components are quite
similar with $z=1.897$. However, the SEDs (in particular, the shapes of
the \ion{C}{4} emission lines; see the lower right panel of
Figure~\ref{fig:qq_pair_spec}) are different, and a deep $I$-band image
does not show a lensing galaxy between the two components. We conclude
that this is a binary quasar rather than a lens. 

\section{Lensed Quasars}\label{sec:dr3_lens}

\begin{figure}
\epsscale{.95}
\plotone{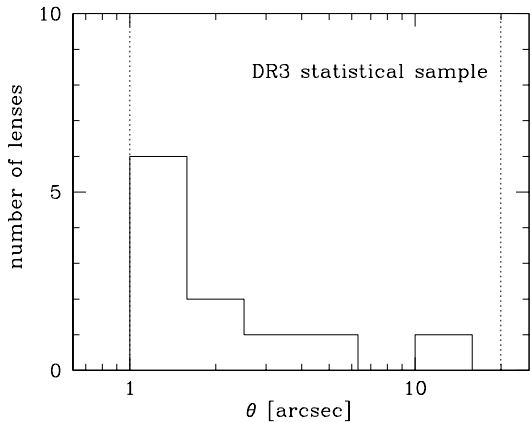}
\caption{Image separation distribution of the SQLS DR3 statistical
 sample in bins of $\Delta\log\theta=0.2$.  The statistical sample is
 constructed in the range $1''<\theta<20''$ as indicated by the dotted
 lines.  The individual lenses are listed in Table~\ref{tab:lens_dr3stat}.
\label{fig:dr3lens}}
\end{figure}

After completing the observations, we have a statistical sample of 11 
lensed quasars with image separations of $1''<\theta<20''$ 
and $i$-band flux ratios (for two-image systems) of faint to bright 
lensed images larger than $10^{-0.5}$. Nine of them are newly discovered 
lensed quasars in the SQLS \citep[e.g.,][]{inada05}, and two of them, 
SDSS~J0913+5259 \citep[SBS~0909+523;][]{oscoz97} and SDSS~1001+5553
\citep[Q0957+561;][]{walsh79}, are previously known lensed quasars. 
The statistical sample is drawn from a subsample (22,683 quasars with 
$0.6 < z < 2.2$ and $i_{\rm cor}>19.1$) of the 46,420 DR3 quasars, 
and therefore some of lenses outside the SDSS DR3
\citep[e.g.,][]{inada06a,inada06b} are not included. Table
\ref{tab:lens_dr3stat} summarizes the DR3 statistical lensed quasar
sample and Figure \ref{fig:dr3lens} shows a histogram of the maximum
image separations of each system. The number of the lenses decreases from
$\theta=1''$ as the image separation increases, consistent with previous
observations \citep[e.g.,][]{browne03}. The statistical sample contains 
9 double lenses and 2 quadruple lenses, and ranges from $\theta=1\farcs0$ 
to $\theta=14\farcs6$, covering both galaxy- and cluster-scale lenses.

There are two lensed quasars listed in Table \ref{tab:can_mor} or 
Table \ref{tab:can_col} that are not part of our statistical sample
in Table \ref{tab:lens_dr3stat}. SDSS~J0832+0404 \citep{oguri07b} 
is a color-selected SDSS lens but it is not included in the 
statistical sample because its $I$-band flux ratio is too extreme. 
The previously known lensed quasar SDSS~J1633+3134 
\citep[FBQ~1633+3134;][]{morgan01} is a morphologically-selected lens,
but its image separation of $\theta=0\farcs66$ is too small to be
included in the statistical sample. These two lenses are listed in
Table~\ref{tab:lens_dr3add}, which contains additional DR3 lensed
quasars outside the statistical sample. 

We also applied our selection algorithms to the DR3 spectroscopic
quasars outside the subsample of 22,683 quasars. For the higher redshift
quasars ($z>2.2$), we used the {\tt star\_L} criteria for the $griz$ 
bands rather than for the $ugri$ bands of the lower redshift lens
selection. We found 4 SDSS lenses, SDSS~J0903+5028 \citep{johnston03}, 
SDSS~J1138+0314 \citep{burles07}, SDSS~J1155+6346 \citep{pindor04},
and SDSS~J1406+6126 \citep{inada07a}, and recovered 2 known 
lenses, SDSS~J0145$-$0945 \citep[Q0142$-$100;][]{reimers02} and 
SDSS~J0911+0550 \citep[RX~J0911+0551;][]{surdej87}. 
Note that SDSS~J0903+5028 was first identified through its compound 
nature (quasar plus luminous red galaxy) of the SDSS spectrum
\citep{johnston03} and SDSS~J1138+0314 was first identified in a 
WB 6.5-m snapshot survey of $\sim 1000$ SDSS quasars \citep{burles07}. 

There are three lenses in the DR3 area which we did not recover as final
follow-up candidate. In two cases, SDSS~J0813+2545
\citep[HS~0810+2554;][]{reimers02} and SDSS~J1650+4251
\citep{morgan03}, the objects are selected in the first stage of our
morphological selection process and then eliminated by the GALFIT
criteria which are designed to select lenses with $\theta>1''$ and
$i$-band flux ratios of faint to bright lensed images
larger than $10^{-0.5}$ (see Paper I). 
SDSS~J0813+2545 (HS~0810+2554) has a small image separation ($<1''$)
and SDSS~J1650+4251 has a small flux ratio ($<10^{-0.5}$), that is why 
they did not pass the GALFIT fitting. The subarcsecond ($0\farcs68$)
radio lens \citep[PMN J0134-0931;][]{winn02,gregg02} was never flagged
at any stage of our survey because of its small image separation. 
These 3 lenses would not be part of our statistical lens sample 
in any case. It does illustrate, however, that our survey is incomplete 
outside of our selection limits.

\section{Summary and Discussions}\label{sec:conc}

We have presented a complete sample of gravitationally lensed quasars
which can be used for various statistical studies. The sample is based
on the 46,420 spectroscopic quasars in the SDSS DR3. We focused on a
subsample of 22,683 quasars with $0.6<z<2.2$ and Galactic
extinction-corrected $i$-band PSF magnitudes brighter than $19.1$. Lens
candidates were identified using the algorithms described here and in
paper I, and verified by extensive follow-up observations in the optical
and near-IR. The resulting 11 lensed quasars constitute a statistical
sample with image separation of $1''<\theta<20''$ and flux
ratios (for two-image systems) of $f_i > 10^{-0.5}$ that should have very 
high completeness based on our tests in paper I. The DR3 spectroscopic 
quasar catalog contains an additional 11 lensed quasars that do not satisfy 
the criteria for our complete sample. Thus we identified a total of 22
lenses, 7 of which were discovered in earlier lens searches other than
the SDSS. 

The lens fraction in our statistical sample, $\sim 0.05\%$, appears
to be lower than in previous studies, but this is largely explained
by our tight criterion on the image separations and flux ratios
and the fact that our sample uses relatively low-redshift faint
quasars. For example, the CLASS survey found a lens fraction of
$\sim 0.14\%$ \citep{browne03}, but $\sim 30\%$ of the lenses have
separations smaller than $1''$ and $\sim 50\%$ of the double
image lenses have flux ratios of faint to bright lensed images
smaller than $10^{-0.5}$. The HST
Snapshot Survey \citep{maoz93} found that $\sim 1\%$ of bright
quasars were lensed, but the magnification bias for bright quasars is 
significantly larger than for our fainter quasar sample. 

This is the first statistical lensed quasar sample that contains both 
galaxy-scale ($\theta\lesssim 3''$) lenses and cluster-scale 
($\theta\gtrsim 10''$) lenses. The maximum image separation in our DR3
statistical sample is $14\farcs62$ (SDSS J1004+4112), much larger than 
the maximum image separation in the CLASS, $4\farcs56$. The distribution
of image separations across the wide mass range from galaxies to
clusters will be valuable in studying the structure formation. We note
that such distributions of splitting angles are obtained from surveys
of lensed galaxies as well. For instance, \citet{cabanac07} presented 40
strongly lensed galaxy candidates, with separations ranging from $2''$
to $15''$. The simple point-like nature of source quasars, however, 
makes it much easier to quantify various selection effects than in 
surveys of lensed galaxies. In addition, our statistical sample 
has an advantage over lensed galaxies (or radio lenses) in
well-understood redshift distribution of the source population. As an
application of the statistical sample, we study cosmological
constraints from the galaxy-scale lenses of the DR3 statistical lens 
sample in Paper III. 

The DR3 spectroscopic quasar sample consists of less than half of the
full SDSS data. Our lens sample will increase significantly in
the future. Assembly of a complete lens sample from the DR5 quasar
catalog \citep{schneider07} is in progress. 

\acknowledgments

N.~I. acknowledges support from the Special Postdoctoral Researcher Program 
of RIKEN and the Japan Society for the Promotion of Science. 
This work was supported in part by Department of Energy contract
DE-AC02-76SF00515.
A portion of this work was also performed under the auspices of the U.S. 
Department of Energy, National Nuclear Security Administration by the
University of California, Lawrence Livermore National Laboratory
under contract No. W-7405-Eng-48. 
M.~A.~S. acknowledges support from NSF grant AST 03-07409.
I.~K. acknowledges supports from Ministry of Education, Culture, Sports, 
Science, and Technology, Grant-in-Aid for Encouragement of Young Scientists 
(No. 17740139), and Grant-in-Aid for Scientific Research on Priority Areas 
No. 467 ``Probing the Dark Energy through an Extremely Wide \& Deep Survey 
with the Subaru Telescope''.
A.~C. acknowledges the support of CONICYT, Chile, under grant FONDECYT
1051061.

Use of the UH 2.2-m telescope and the UKIRT 3.8-m telescope for the
observations is supported by the National Astronomical
Observatory of Japan (NAOJ). 
Based in part on observations obtained with the Apache Point Observatory
3.5-meter telescope, which is owned and operated by the Astrophysical
Research Consortium. 
Based in part on data collected at Subaru Telescope (some of data
obtained from the Subaru Telescope Sciences Archive System [SMOKA]), 
which is operated by NAOJ. 
Some of the data presented herein were obtained at the W.M. Keck
Observatory, which is operated as a scientific partnership among the
California Institute of Technology, the University of California and the
National Aeronautics and Space Administration. The Keck Observatory was 
made possible by the generous financial support of the W.M. Keck
Foundation. 
The WIYN Observatory is a joint facility of the University of
Wisconsin-Madison, Indiana University, Yale University, and the
National Optical Astronomy Observatories. 
This work is also based in part on observations obtained with the MDM
2.4m Hiltner telescope, which is owned and
operated by a consortium consisting of Columbia University, Dartmouth
College, the University of Michigan, the Ohio State University and
Ohio University. 
The WB 6.5-m telescope is the first telescope of
the Magellan Project; a collaboration between the Observatories of
the Carnegie Institution of Washington, University of
Arizona, Harvard University, University of Michigan, and
Massachusetts Institute of Technology to construct two 6.5 Meter
optical telescopes in the southern hemisphere.
Based in part on observations made with the NASA/ESA Hubble Space 
Telescope, obtained at the Space Telescope Science Institute, 
which is operated by the Association of Universities for Research 
in Astronomy, Inc., under NASA contract NAS 5-26555. These 
observations are associated with HST program GO-9744. 
Based in part on observations made with telescopes (ESO 3.6-m and NTT) 
at the European Southern Observatories La Silla in Chile.
Some observations reported here were obtained at the MMT Observatory, 
a joint facility of the University of Arizona and the Smithsonian 
Institution. 

Funding for the SDSS and SDSS-II has been provided by the 
Alfred P. Sloan Foundation, the Participating Institutions, 
the National Science Foundation, the U.S. Department of Energy, 
the National Aeronautics and Space Administration, the Japanese 
Monbukagakusho, the Max Planck Society, and the Higher Education 
Funding Council for England. The SDSS Web Site is http://www.sdss.org/.

The SDSS is managed by the Astrophysical Research Consortium for 
the Participating Institutions. The Participating Institutions 
are the American Museum of Natural History, Astrophysical Institute 
Potsdam, University of Basel, Cambridge University, Case Western 
Reserve University, University of Chicago, Drexel University, 
Fermilab, the Institute for Advanced Study, the Japan Participation 
Group, Johns Hopkins University, the Joint Institute for Nuclear 
Astrophysics, the Kavli Institute for Particle Astrophysics and 
Cosmology, the Korean Scientist Group, the Chinese Academy of 
Sciences (LAMOST), Los Alamos National Laboratory, the Max-Planck-Institute 
for Astronomy (MPIA), the Max-Planck-Institute for Astrophysics (MPA), 
New Mexico State University, Ohio State University, University of 
Pittsburgh, University of Portsmouth, Princeton University, the 
United States Naval Observatory, and the University of Washington.

\clearpage

\LongTables

\begin{deluxetable}{lccccllll}
\tabletypesize{\footnotesize}
\tablecaption{MORPHOLOGICAL CANDIDATES\label{tab:can_mor}}
\tablewidth{0pt}
\tablehead{ 
\colhead{Object} & 
\colhead{$z$\tablenotemark{a}} &
\colhead{$i_{\rm cor}$\tablenotemark{b}} & 
\colhead{$\theta_{\rm SDSS}$\tablenotemark{c}} & 
\colhead{${\Delta}{i}$\tablenotemark{c}} & 
\colhead{image\tablenotemark{d}} & 
\colhead{spec\tablenotemark{d}} & 
\colhead{comment} & 
\colhead{Ref.} 
}
\startdata
SDSS~J001125.75$-$105710.4  &  0.641  &  18.95  &  0.48  &  0.70  &  UF($K$)          &  \nodata\phn  &  single QSO  &  \nodata\phn  \\
SDSS~J004242.00+135450.0    &  1.622  &  16.72  &  0.56  &  0.81  &  UF($K$)          &  \nodata\phn  &  single QSO  &  \nodata\phn  \\
SDSS~J011229.41+151213.9    &  1.957  &  18.86  &  1.12  &  0.30  &  WF($i$)          &  \nodata\phn  &  QSO+galaxy  &  \nodata\phn  \\
SDSS~J012259.49+151147.2    &  1.276  &  18.17  &  0.45  &  0.36  &  UF($K$)          &  \nodata\phn  &  single QSO  &  \nodata\phn  \\
SDSS~J020707.59$-$100541.6  &  0.662  &  18.78  &  0.48  &  0.10  &  UF($K$)          &  \nodata\phn  &  single QSO  &  \nodata\phn  \\
SDSS~J021249.59+003448.7    &  1.222  &  18.86  &  1.69  &  0.73  &  8k($I$),WF($i$)  &  WF           &  QSO+galaxy  &  \nodata\phn  \\
SDSS~J021645.80$-$010204.8  &  1.087  &  18.55  &  0.53  &  1.35  &  UF($K$)          &  \nodata\phn  &  single QSO  &  \nodata\phn  \\
SDSS~J024634.09$-$082536.1  &  1.685  &  17.77  &  1.03  &  1.10  &  AC($V$,$I$),Ma($i$),NR($K'$),NM($H$) &  ES      &  SDSS Lens  &  1  \\
\enddata
\tablenotetext{a}{Redshifts from the SDSS DR3 quasar catalog} 
\tablenotetext{b}{$i$-band PSF magnitudes with Galactic extinction
 corrections from the SDSS DR3 quasar catalog}  
\tablenotetext{c}{Image separations ($\theta_{\rm SDSS}$) in units of
 arcsec and magnitude differences (${\Delta}{i}$) between the expected
 two components, derived from fitting the SDSS $i$-band image with two PSFs 
 using GALFIT.} 
\tablenotetext{d}{Instruments (and filters) used for the follow-up observations. 
                  8k: UH8k at UH88, QU: QUIRC at UH88, Op: Optic at
		      UH88, WF: WFGS2 at UH88, Te: Tek2048 at UH88,  
                  DA: DIS III at ARC 3.5m, 
                  ES: ESI at Keck, LR: LRIS at Keck, NR: NIRC at Keck, 
                  UF: UFTI at UKIRT, 
                  FO: FOCAS at Subaru, SC: Suprime-Cam at Subaru,
                  Ma: MagIC at WB 6.5m, DW: Double Imaging Spectrograph at WB 6.5m,
                  AC: ACS at HST, NM: NICMOS at HST, 
                  RE: RETROCAM at MDM 2.4m, 
                  MS: MMT spectrograph, AR: ARIES at MMT, 
                  EF: EFOSC2 at ESO 3.6m,
                  EM: EMMI at NTT,
                  MM: MiniMo at WIYN.}
\tablerefs{(1) \citealt{inada05}; (2) \citealt{pindor06}; 
(3) \citealt{oscoz97}; (4) \citealt{inada03a}; (5) \citealt{hennawi06}; 
(6) \citealt{inada07b}; (7) \citealt{morokuma07}; (8) \citealt{oguri04b}; 
(9) \citealt{oguri07b}; (10) \citealt{morgan01}.}
\end{deluxetable}


\begin{deluxetable}{lccrllll}
\tabletypesize{\footnotesize}
\tablecaption{COLOR CANDIDATES\label{tab:can_col}}
\tablewidth{0pt}
\tablehead{ 
\colhead{Object} & 
\colhead{$z$\tablenotemark{a}} &
\colhead{$i_{\rm cor}$\tablenotemark{b}} & 
\colhead{$\theta_{\rm SDSS}$\tablenotemark{c}} & 
\colhead{image\tablenotemark{d}} & 
\colhead{spec\tablenotemark{d}} & 
\colhead{comment} & 
\colhead{Ref.} 
}
\startdata
SDSS~J004757.25+144741.9    &  1.612        &  18.67  &         &               &               &                              &               \\
SDSS~J004757.87+144744.7    &  (2.790)      &  18.72  &   9.42  &  \nodata\phn  &  DA           &  QSO pair                    &  \nodata\phn  \\
\hline
SDSS~J021649.25$-$003723.5  &  1.542        &  18.55  &         &               &               &                              &               \\
SDSS~J021649.15$-$003711.5  &  \nodata\phn  &  19.57  &  12.15  &  \nodata\phn  &  DA           &  QSO+star                    &  \nodata\phn  \\
\hline
SDSS~J023205.09+010640.2    &  1.259        &  17.28  &         &               &               &                              &               \\
SDSS~J023205.18+010634.2    &  \nodata\phn  &  17.90  &   6.20  &  \nodata\phn  &  DA           &  QSO+star                    &  \nodata\phn  \\
\hline
SDSS~J025804.27$-$001059.9  &  1.551        &  18.93  &         &               &               &                              &               \\
SDSS~J025803.81$-$001118.1  &  \nodata\phn  &  20.05  &  19.43  &  \nodata\phn  &  DA           &  different SED               &  \nodata\phn  \\
\enddata
\tablecomments{Two candidates that are identified by the morphological
  selection algorithm as well, SDSS~J073406.75+273355.6 and
  SDSS~J133534.79+011805.5, are listed in Table \ref{tab:can_mor}.}
\tablenotetext{a}{Redshifts from the SDSS DR3 quasar catalog and the
  follow-up observations (in parentheses) }  
\tablenotetext{b}{$i$-band PSF magnitudes with the Galactic extinction
  correction from the SDSS DR3 quasar catalog}  
\tablenotetext{c}{Image separations in units of arcsec between two
  components from the SDSS imaging data.} 
\tablenotetext{d}{Instruments (and filters) used for the follow-up
                  observations.  
                  8k: UH8k at UH88, QU: QUIRC at UH88, Op: Optic at UH88, WF: WFGS2 at UH88, Te: Tek2048 at UH88, 
                  DA: DIS III at ARC 3.5m, SP: SPIcam at ARC 3.5m, 
                  LR: LRIS at Keck, NR: NIRC at Keck, 
                  UF: UFTI at UKIRT,
                  SC: Suprime-Cam at Subaru,
                  MS: MMT spectrograph, 
                  EF: EFOSC2 at ESO 3.6m.
                  }
\tablerefs{(1) \citealt{hennawi06}; 
           (2) \citealt{oguri07b};
           (3) \citealt{oguri05}; 
           (4) \citealt{walsh79}; 
           (5) \citealt{inada03b};
           (6) \citealt{oguri04a}; 
           (7) \citealt{croom04}; 
           (8) \citealt{pindor06}.}
\end{deluxetable}


\begin{deluxetable}{lccrcll}
\tabletypesize{\footnotesize}
\tablecaption{LENSED QUASARS FROM THE SDSS DR3: STATISTICAL SAMPLE \label{tab:lens_dr3stat}}
\tablewidth{0pt}
\tablehead{ 
\colhead{Object} & 
\colhead{$N_{\rm img}$} & 
\colhead{$z$\tablenotemark{a}} &
\colhead{$\theta_{\rm max}$\tablenotemark{b}} & 
\colhead{$f_{i}$ or $f_{I}$ \tablenotemark{c}} & 
\colhead{Comment} & 
\colhead{Ref.} 
}
\startdata
SDSS~J0246$-$0825 & 2 & 1.682  &   1.04  &  0.34  &  SDSS lens                &  1  \\ 
SDSS~J0913+5259   & 2 & 1.377  &   1.14  &  0.70  &  known lens SBS~0909+523  &  2, 3 \\
SDSS~J0924+0219   & 4 & 1.524  &   1.78  &  0.43  &  SDSS lens                &  4  \\ 
SDSS~J1001+5027   & 2 & 1.838  &   2.86  &  0.72  &  SDSS lens                &  5  \\ 
SDSS~J1001+5553   & 2 & 1.405  &   6.17  &  0.94  &  known lens Q0957+561     &  3, 6\\
SDSS~J1004+4112   & 5 & 1.732  &  14.62  &  0.23  &  SDSS lens                &  7, 8  \\ 
SDSS~J1021+4913   & 2 & 1.720  &   1.14  &  0.40  &  SDSS lens                &  9  \\ 
SDSS~J1226$-$0006 & 2 & 1.121  &   1.24  &  0.45  &  SDSS lens                &  10 \\
\enddata
\tablenotetext{a}{Redshifts from the follow-up observations} 
\tablenotetext{b}{Maximum image separations in units of arcsec.}
\tablenotetext{c}{Flux ratios between the brightest lensed image and the 
                  farthest lensed image from the brightest image, in the $I$- or 
                  $i$-band images.}
\tablerefs{(1) \citealt{inada05}; 
           (2) \citealt{oscoz97}; 
           (3) CASTLES webpage (C.~S.~Kochanek et al., http://cfa-www.harvard.edu/castles/.);
           (4) \citealt{inada03a}; 
           (5) \citealt{oguri05}; 
           (6) \citealt{walsh79};
           (7) \citealt{inada03b}; 
           (8) \citealt{oguri04a}; 
           (9) \citealt{pindor06};
           (10) \citealt{inada07b}; 
           (11) \citealt{morokuma07}; 
           (12) \citet{oguri04b};
           (13) \citealt{oguri07b}.}
\end{deluxetable}


\begin{deluxetable}{lccrclll}
\tabletypesize{\footnotesize}
\tablecaption{ADDITIONAL LENSED QUASARS IN THE SDSS DR3 QUASAR CATALOG\label{tab:lens_dr3add}}
\tablewidth{0pt}
\tablehead{ 
\colhead{Object} & 
\colhead{$N_{\rm img}$} & 
\colhead{$z$\tablenotemark{a}} &
\colhead{$\theta$\tablenotemark{b}} & 
\colhead{$f_{i}$ or $f_{I}$ \tablenotemark{c}} & 
\colhead{Comment} & 
\colhead{Reject\tablenotemark{d}} & 
\colhead{Ref.} 
}
\startdata
SDSS~J0134$-$0931 & 5 &  2.216  &   0.68  &  0.03  &  known lens PMN~J0134$-$0931  &  too small $\theta$             &  1, 2, 3  \\  
SDSS~J0145$-$0945 & 2 &  2.719  &   2.23  &  0.15  &  known lens Q0142$-$100       &  $z>2.2$, $f_I<10^{-0.5}$       &  3, 4     \\  
SDSS~J0813+2545   & 4 &  1.500  &   0.91  &  0.06  &  known lens HS~0810+2554      &  by GALFIT fitting              &  3, 5     \\ 
SDSS~J0832+0404   & 2 &  1.115  &   1.98  &  0.22  &  SDSS lens                    &  $f_I<10^{-0.5}$                &  6        \\ 
SDSS~J0903+5028   & 2 &  3.584  &   2.80  &  0.46  &  SDSS lens                    &  $z>2.2$                        &  7        \\
SDSS~J0911+0550   & 4 &  2.800  &   3.26  &  0.41  &  known lens RX~J0911+0551     &  $z>2.2$                        &  3, 8     \\
SDSS~J1138+0314   & 4 &  2.442  &   1.44  &  0.35  &  SDSS lens                    &  $z>2.2$                        &  3, 9     \\
SDSS~J1155+6346   & 2 &  2.890  &   1.83  &  0.50  &  SDSS lens                    &  $z>2.2$                        &  3, 10    \\
\enddata
\tablecomments{See text for the selection of each lensed quasar.} 
\tablenotetext{a}{Redshifts from  follow-up observations.} 
\tablenotetext{b}{Maximum image separations in units of arcsec.}
\tablenotetext{c}{Flux ratios between the brightest lensed image and the 
                  farthest lensed image from the brightest image, in the $I$- or 
                  $i$-band images.} 
\tablenotetext{d}{The reason that each lens is excluded from the
  statistical sample.} 
\tablerefs{ (1) \citealt{winn02}; 
            (2) \citealt{gregg02}; 
            (3) CASTLES webpage (C.~S.~Kochanek et al., http://cfa-www.harvard.edu/castles/.);
            (4) \citealt{surdej87}; 
            (5) \citealt{reimers02}; 
            (6) \citealt{oguri07b};
            (7) \citealt{johnston03};
            (8) \citealt{bade97}; 
            (9) \citealt{burles07}; 
            (10) \citealt{pindor04}; 
            (11) \citealt{inada07a}; 
            (12) \citealt{morgan01};
            (13) \citealt{morgan03}.}
\end{deluxetable}

\end{document}